\documentclass[usenatbib]{mn2e}
\usepackage{epsfig}
\usepackage{amsmath}
\usepackage{ulem}

\def\gsim{\mathrel{\raise0.35ex\hbox{$\scriptstyle >$}\kern-0.6em 
\lower0.40ex\hbox{{$\scriptstyle \sim$}}}}
\def\lsim{\mathrel{\raise0.35ex\hbox{$\scriptstyle <$}\kern-0.6em 
\lower0.40ex\hbox{{$\scriptstyle \sim$}}}}
\def\gs{\mathrel{\raise0.35ex\hbox{$\scriptstyle >$}\kern-0.6em 
\lower0.40ex\hbox{{$\scriptstyle \sim$}}}}
\def\ls{\mathrel{\raise0.35ex\hbox{$\scriptstyle <$}\kern-0.6em 
\lower0.40ex\hbox{{$\scriptstyle \sim$}}}}

\def\kms {{\,\rm km\,s^{-1}}}

\def\lesssim{\mathrel{\hbox{\rlap{\hbox{\lower4pt\hbox{$\sim$}}}\hbox{$<$}}}}
\def\gtrsim{\mathrel{\hbox{\rlap{\hbox{\lower4pt\hbox{$\sim$}}}\hbox{$>$}}}}

\date{\today}
\title[A Deficit of Faint Red Galaxies]
{A Deficit of Faint Red Galaxies in the Possible Large-Scale Structures
around the RDCS J1252.9--2927 Cluster at $z=1.24$}

\author[M. Tanaka et al.]{
\parbox[t]{\textwidth}{
Masayuki Tanaka$^1$,
Tadayuki Kodama$^{2}$,
Masaru Kajisawa$^2$,
Richard Bower$^3$,
Ricardo Demarco$^4$,
Alexis Finoguenov$^5$,
Chris Lidman$^6$,
Piero Rosati$^7$,
}
\vspace*{6pt}\\
$^{1}$Department of Astronomy, School of Science, University of Tokyo, Tokyo 113--0033, Japan \\
$^{2}$National Astronomical Observatory of Japan, Mitaka, Tokyo 181--8588, Japan \\
$^{3}$Department of Physics, University of Durham, South Road, Durham DH1 3LE, UK \\
$^{4}$Department of Physics and Astronomy, The Johns Hopkins University, 3400 North Charles Street, Baltimore, MD 21218-2686, USA \\
$^{5}$Max-Planck-Institut f\"{u}r extraterrestrische Physik, Giessenbachstrasse, 85748 Garching, Germany\\
$^{6}$European Southern Observatory, Alonso de Cordova 3107, Vitacura, Casilla 19001, Santiago 19, Chile\\
$^{7}$European Southern Observatory, Karl Schwarzschild Strasse 2, Garching bei Muenchen, D-85748, Germany
}

\begin{document}

\maketitle

\begin{abstract}
We report a discovery of possible large-scale structures around the RDCS J1252.9$-$2927 cluster
at $z=1.24$ based on photometric redshifts.
We carried out multi-band wide-field imaging with Suprime-Cam on the Subaru Telescope
and WFCAM on the United Kingdom Infra-Red Telescope (UKIRT).
Our data cover an approximately $35'\times30'$ field in $V$, $R$, $i'$, $z'$, and $K$ bands.
We apply a photometric redshift technique to extract galaxies at or near the cluster redshift.
The distribution of photo-$z$ selected galaxies reveals clumpy structures
surrounding the central cluster.
It seems that there is a large ($>20$ Mpc) filamentary structure
extending in the NE-SW direction.
We compare the observed structure with an X-ray map and find that
two of the four plausible clumps show significant X-ray emissions
and one with a marginal detection,
which strongly suggest that they are dynamically bound systems.
Following the discovery of the possible large-scale structure, we carried  out
deeper SOFI $K_s$-band imaging with New Technology Telescope on four plausible clumps.
We construct the optical-to-near-infrared colour-magnitude diagrams of the galaxies
in the clumps, and find that the colour-magnitude relation (CMR) of
the red galaxies in the clumps is sharply truncated below $K_s=22$.
Few faint red galaxies are seen in these clumps.
This suggests that the CMR first appears at the bright magnitudes and it extends to the
fainter magnitudes with time, which is consistent with the 'down-sizing' picture.
Interestingly, the CMR of the main cluster has previously been shown to have
a clear relation down to $K_s=23$ \citep{lidman04}.
This confirms our previous suggestion that the build-up of the CMR
is delayed in low-density environments.
All in all, we suggest that galaxies follow the 'environment-dependent down-sizing' evolution.
Massive galaxies in high density environments first stop forming stars and become red.
Less massive galaxies in less dense environments become red at later times.
Based on a few assumptions, we predict that the brightest tip of the
CMR appears at $z\sim2.5$.
\end{abstract}


%
%
\section{Introduction}

\label{sec:intro}

Formation and evolution of structures in the Universe has been
one of the central issues in modern astronomy.
Filamentary structures are predicted to grow from the initial density fluctuations
of the Universe and filaments are distributed all over the Universe forming the cosmic web.
At crossroads of filaments, we often see massive clusters of galaxies
(e.g. RXJ0152.7$-$13 cluster; \citealt{demarco05,girardi05,maughan06,tanaka06}).
Clusters should grow through the hierarchical assembly of galaxies flowing along
filaments, and indeed they are often found to accompany filamentary/clumpy structures
around them (e.g., \citealt{kodama01,kodama05,gal04,tanaka06}).
In this paper, we focus on one of the highest redshift X-ray clusters to date,
RDCS J1252.9--2927 (RDCSJ1252 hereafter) at $z=1.24$ \citep{rosati04}.

RDCSJ1252 was discovered in {\it ROSAT Deep Cluster Survey} and 
the mass of the cluster was estimated to be
${\rm M(<0.5Mpc)}\sim2\times10^{14} {\rm M_\odot}$ \citep{rosati04}.
Following the discovery, the cluster has been a subject of extensive observations.
\citet{blakeslee03} reported ACS imaging on the cluster and presented
a tight colour-magnitude relation (CMR) of early-type galaxies.
The superb resolution of ACS enabled a clear detection of the weak shear signal
from this high-$z$ cluster \citep{lombardi05}.
In addition to the optical HST imaging, ground-based near-IR imaging
was also conducted.
\citet{lidman04} showed a clear near-IR CMR, and 
\citet{toft04} claimed that luminosity function of the cluster galaxies
has a shallower slope than that of a local cluster.
Recently, \citet{strazzullo06} derived a $K_s$ band LF of
the cluster based on deep near-IR imaging
(they also derived LFs of two other $z\sim1$ clusters).
An analysis of the Fundamental plane was also performed based on the deep
spectroscopy \citep{holden05}.
We extend the investigation of this cluster further by obtaining
the new and unique multi-band wide-field images of this cluster.

In this paper, we focus on the faint end of the CMR of red galaxies.
Recent work on CMR suggested that high redshift clusters exhibit
a deficit of faint red galaxies compared to local clusters
\citep{kajisawa00,nakata01,kodama04,delucia04,tanaka05,delucia06,homeier06}.
This is consistent with the 'down-sizing' in star formation activities
of galaxies \citep{cowie96}  --- hosts of active star formation
shifts from massive galaxies to less massive ones as the Universe ages.
It seems that this down-sizing depends on large-scale environment
in the sense that it is delayed in low-density environments
(\citealt{tanaka05}; but see also \citealt{delucia06}).
In this paper,
we put an emphasis on the environmental variation of the faint end
of the CMR to test this 'environment-dependent down-sizing' picture.

The outline of this paper is as follows.
In Section 2, we describe our observations and data reduction.
Then we present distribution of galaxies around RDCSJ1252 in Section 3.
Section 4 describes deep SOFI observations of the discovered clumps of galaxies.
We examine the CMRs of red galaxies in the structure in Section 5.
Finally, we discuss and summarize our findings in Sections 6 and 7.
Throughout this paper, we assume a flat Universe with
$\Omega_{\rm M}=0.3,\ \Omega_{\rm \Lambda}=0.7$ and H$_0=70\kms \rm Mpc^{-1}$.
All the magnitudes are presented on the AB system.
We use the following abbreviations: CMD for colour-magnitude diagram, 
CMR for colour-magnitude relation, and LF for luminosity function.

%
%
\section{Subaru and UKIRT Observations}
\label{sec:obs}
We observed RDCSJ1252 with Suprime-Cam \citep{miyazaki02} on the Subaru Telescope 
in the $VRi'z'$ bands under photometric conditions.
The images are smoothed to their worst seeing size of $\sim0.8$ arcsec in FWHM.
Since this cluster lies at a high redshift, even the $z'$-band does not firmly straddle
the 4000$\rm\AA$ break, and a good performance of photometric redshift cannot be achieved
with the optical data only.
Therefore, we carried out $K$-band imaging with WFCAM on UKIRT \citep{henry00}.
The sky was mostly photometric throughout the run.
However, the seeing condition was poor and the FWHM of stellar images was
$\sim1.3$ arcsec.
Reduction of all the data was performed in a standard manner.
Table \ref{tab:obs} summarizes the observation dates,
total exposure times and limiting magnitudes.
The limiting magnitude is evaluated as a $5\sigma$ limit in 2 arcsec apertures,
which is derived from a flux distribution within 2 arcsec apertures randomly placed on each image.
Photometric zero-points were determined from standard stars observed during the run
($VRi'z'$) and from the 2MASS data ($K$; \citealt{jarrett00}).
We shifted the zero-points of the optical data ($\sim 0.05$ mag.)
so that colours of the stars in our field match those from \citet{gunn83}.
Galactic extinction is corrected using the dust map from \citet{schlegel98}.

We use {\scriptsize SEXTRACTOR} \citep{bertin96} to extract
galaxies from our combined images.
Object detection is performed in the $z'$-band, which is much cleaner
(much less artifacts) and slightly deeper than the $K$-band image for red galaxies at $z=1.2$.
We then limit the catalogue to $K<22.4$ (5$\sigma$) to approximate
a stellar-mass limited sample.
Star-galaxy separation is performed on the  $R-z'$  vs. $z'-K$ diagram.
We do not smooth the optical images to match the relatively bad seeing sizes
in the UKIRT data.
Instead, we use Kron magnitudes ({\sc MAG\_AUTO}) for both total magnitudes
and colours, rather than those measured in a fixed aperture size.

\begin{table}
\caption{
Observation log.  The limiting magnitudes ($M_{lim}$) are evaluated as
a 5$\sigma$ limit in 2 arcsec aperture.
The exposures are not uniform across the WFCAM and SOFI fields,
and the values quoted below are averages.
}
\label{tab:obs}
\begin{tabular}{ccccc}
\hline
Filter & Instrument & Date              & Exp. (min) & $M_{lim}$\\
\hline
$V$   & Suprime-Cam &      6 May 2005   & 60  & 26.4\\
$R$   & Suprime-Cam &      5 May 2005   & 80  & 26.4\\
$i'$  & Suprime-Cam &      6 May 2005   & 89  & 25.9\\
$z'$  & Suprime-Cam &      5 May 2005   & 55  & 25.0\\
$K$   & WFCAM       & 24--27 May 2005   & 97  & 22.4\\
$K_s$ & SOFI        & 20--23 March 2006 & 164 & 23.0\\
\hline
\end{tabular}
\end{table}

%
%
\section{Large-Scale Structures}
\label{sec:lss}
\subsection{Photometric Identification}

The cluster lies at a high redshift of $z=1.237$ (Demarco et al. 2007, submitted)
and 
 galaxies at different redshifts (foreground galaxies in particular) are
significant contamination when examining galaxies at the cluster redshift.
To reduce the contamination, we apply the photometric redshift technique 
taking advantage of the multi-band data ($VRi'z'K$).
We use the photometric redshift code of \citet{kodama99}.
Star formation histories of the model templates are described by a
combination of an elliptical galaxy model and a disc model (see \citealt{kodama99} for details).
\citet{lidman04} examined the CMR of the RDCSJ1252 cluster and found
that the CMR of cluster galaxies is fitted with the passive evolution model
with a formation redshift of $2<z_f<3$.
Motivated by this, we set $z_f=2.5$ in our models and run the photometric redshift code.

\begin{figure}
\begin{center}
\leavevmode
\epsfxsize 0.9\hsize \epsfbox{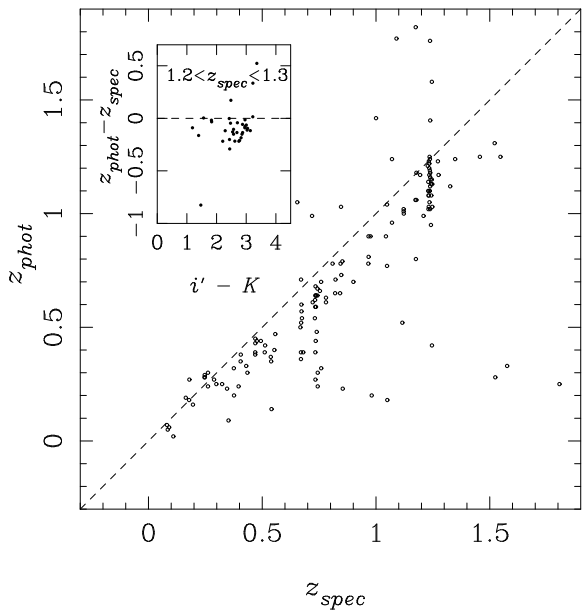}
\end{center}
\caption{
Photometric redshift plotted against spectroscopic redshift
(Demarco et al. submitted.).
In the inset, differences between photometric and spectroscopic
redshifts are plotted as a function of $i'-K$ colour using the 
galaxies at $1.2<z_{spec}<1.3$.
}
\label{fig:photoz_accuracy}
\end{figure}

Fig. \ref{fig:photoz_accuracy} shows the accuracy of our photometric redshifts.
Spectroscopic redshifts are taken from Demarco et al. (submitted).
We see a small systematic offset ($z_{phot}-z_{spec}=-0.1$), but it does not
affect the analyses presented below.
Although some catastrophic failures can be seen, overall correlation is relatively good.
Using galaxies at $1.2<z_{spec}<1.3$, we find that the median of
$z_{phot}-z_{spec}$ is $-0.10$ and the dispersion around it is $0.07$ excluding
the obvious outliers ($|z_{phot}-z_{spec}+0.1|>0.2$).
The fraction of the catastrophic failures at $1.2<z_{spec}<1.3$ is 11 per cent ($4/35$).
Roughly 80 per cent ($34/43$) of $1.0<z_{phot}<1.3$ galaxies lie at $1.14<z_{spec}<1.34$
(i.e., $|z_{spec}-z_{cl}|<0.1$, where $z_{cl}$ is the cluster redshift).
We do not see strong colour dependence of the photo-$z$ accuracy
contrary to our expectations \citep{tanaka06},
although this could be simply because we have few blue galaxies
with spectroscopic redshifts.
We compare our photometric redshifts with those estimated with 
the publicly available photometric redshift code {\sc hyperz} \citep{bolzonella00}.
We find that {\sc hyperz} gives a poorer accuracy ---
based on galaxies at $1.2<z_{spec}<1.3$, the median of
$z_{phot}-z_{spec}$ is $-0.11$ and the dispersion around it is $0.15$
excluding outliers.  The fraction of the outliers is similar to ours ($5/35$).

An effect of object blending on photometry may be significant
in high-density regions, where objects often overlap with one another.
Due to this effect, we may not be able to accurately measure colours of galaxies
in high-density environments, which may result in poor photo-$z$ accuracy.
However, we estimate that 18 per cent ($9/51$) of strongly blended objects show
catastrophic failure ($|z_{phot}-z_{spec}+0.1|>0.2$),
while 19 ($18/97$) per cent of unblended objects show such failures.
Thus, the object blending does not seem to affect our analysis
in a significant way.

We present in Fig. \ref{fig:lss} the smoothed projected density distribution
of photo-$z$ selected galaxies.
Here we adopt a redshift slice of $1.0<z_{phot}<1.3$.
As described above, 80 per cent  of the spectroscopic objects in this photo-$z$ range
fall in $|z_{spec}-z_{cl}|<0.1$.
The remaining 20 per cent spread over a wide redshift range, and they will not
be a source of spurious structures.

The RDCSJ1252 cluster can be clearly identified at the centre of the field.
Interestingly, we see a chain of clumps towards the NNE direction
from the cluster extending over 15 Mpc in comoving scale.
There are a few other clumps located at ENE and WSW of the main cluster.
All of these clumps show the density contrast greater than $5\sigma$.
No such density contrast is seen either at NW or at SE of the main cluster,
except the two at the SE edge of our field of view
which we do not examine in what follows due to the close proximity to the field edge.
We note that the structures do not strongly change if we adopt
a larger smoothing scale of galaxy density (e.g. 1 Mpc in comoving),
although the significance of the compact clumps slightly decreases.
As a further check, we apply adaptive smoothing with {\sc csmooth}
in the {\sc ciao} package developed for Chandra X-ray data and confirm
that the structure seen in Fig. \ref{fig:lss} is reproduced.
Even if we apply colour-cuts around the red sequence of
cluster members (in $Ri'K$) instead of the photo-$z$ selection,
most of the structures (except for a few clumps at SW) are reproduced.
To see if the observed structure is unique to $z_{phot}\sim1.2$,
we place photometric redshift slices at different
redshifts and perform the same analysis as above.
The structure is not observed at different redshifts at a significant level and
found to be unique to $z_{phot}\sim1.2$,
except that we see an over-density
of galaxies around the main cluster at $0.4<z_{phot}<0.7$.
This must be a foreground system at $z\sim0.7$ as is also evident in Fig. \ref{fig:photoz_accuracy}.
The accuracy of the photometric redshifts may still remain a concern.
However, galaxies with wrong photometric redshifts would only weaken
the significance of any true structures and would not produce any
artificial structures.

\begin{figure*}
\begin{center}
\leavevmode
\epsfxsize 0.33\hsize \epsfbox{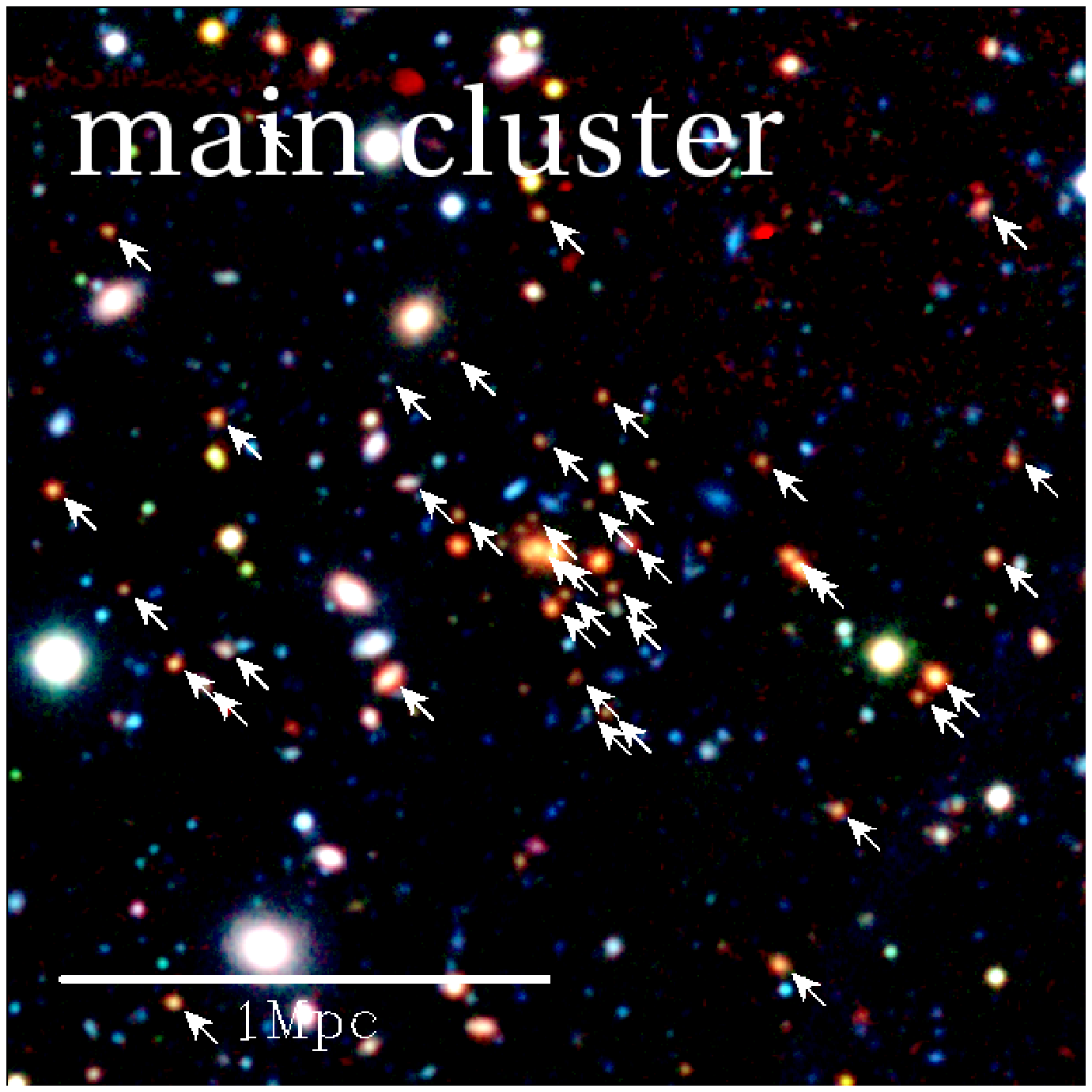}
\epsfxsize 0.33\hsize \epsfbox{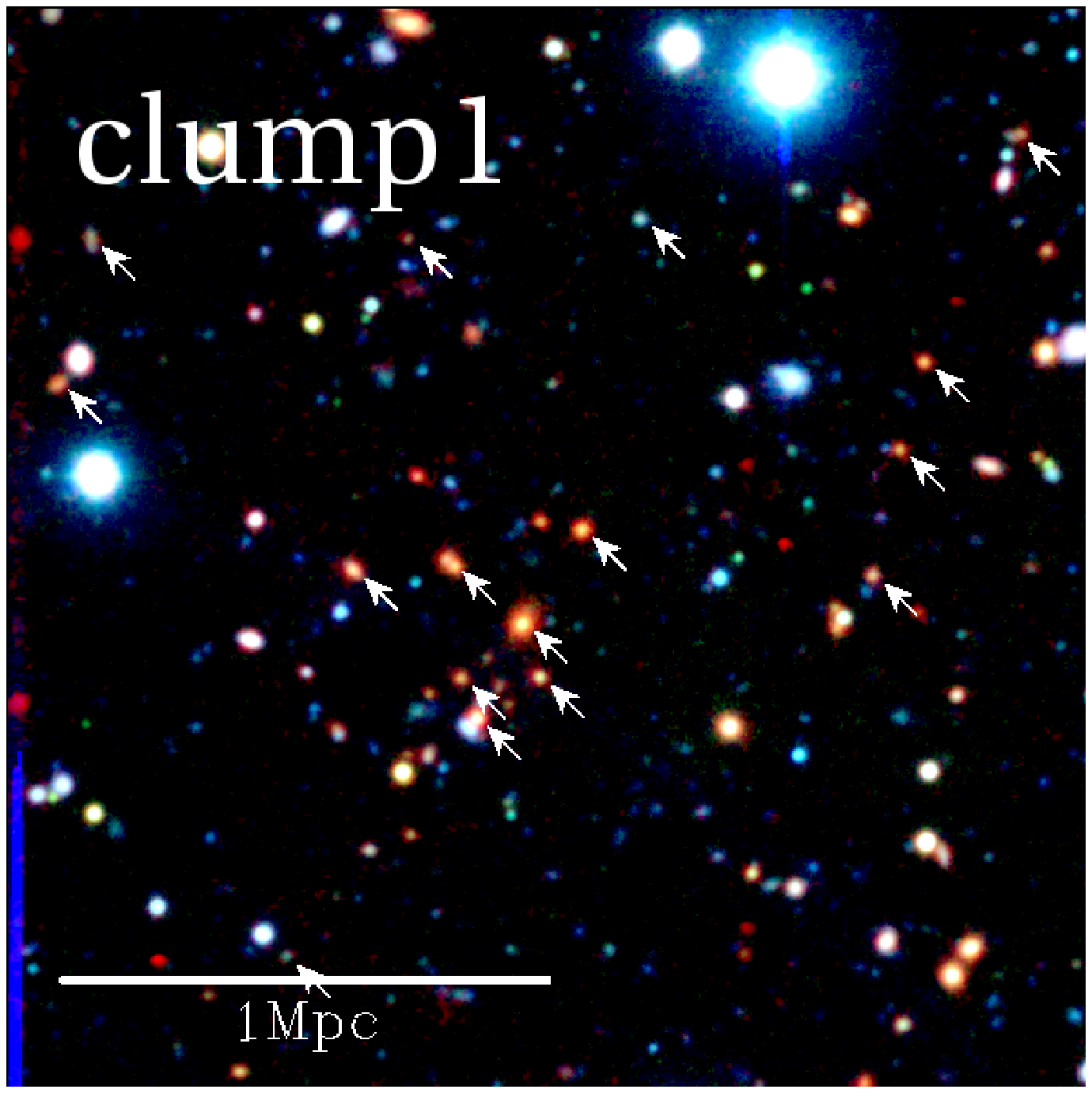}
\epsfxsize 0.33\hsize \epsfbox{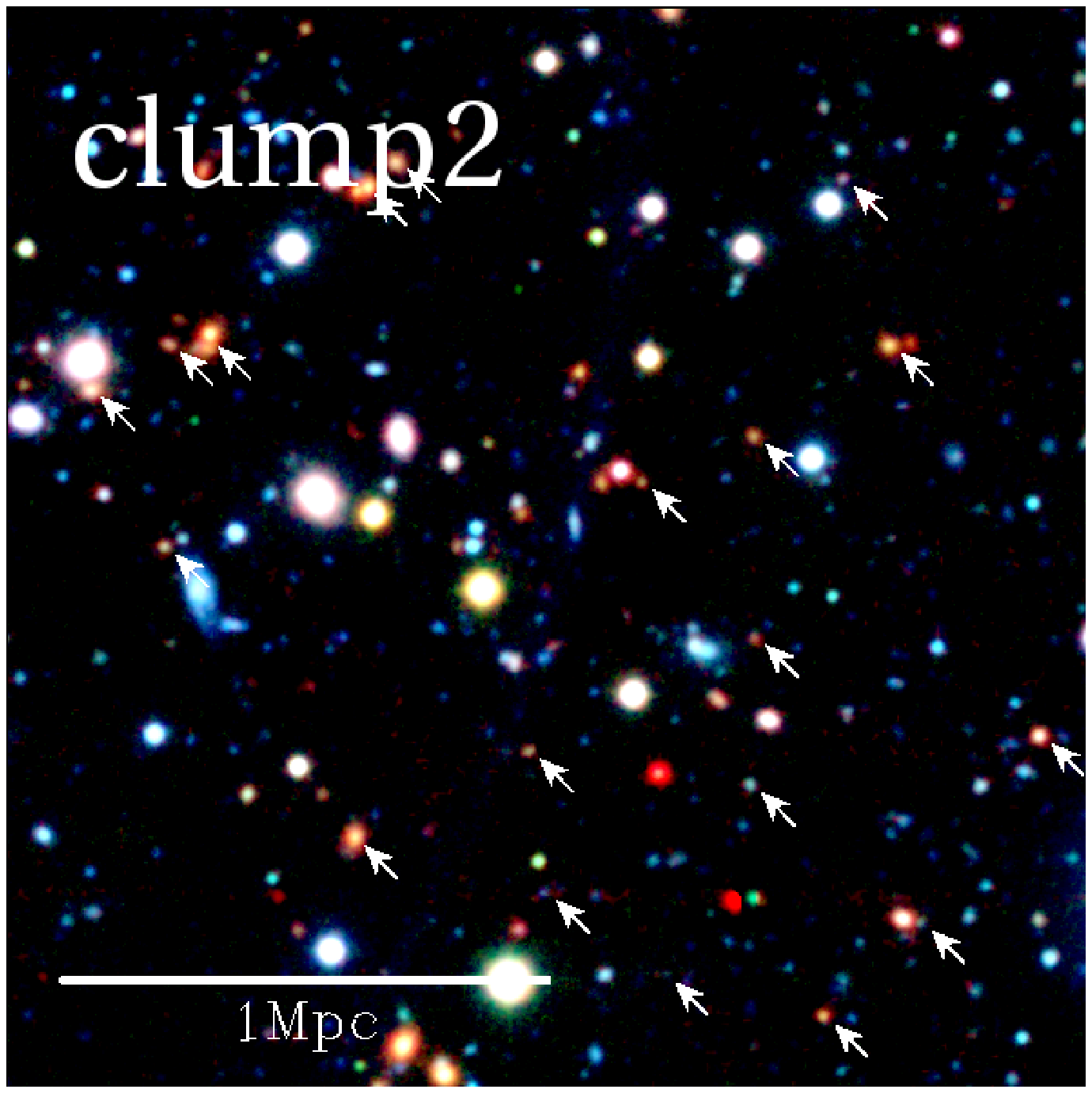}\\
\epsfxsize 0.33\hsize \epsfbox{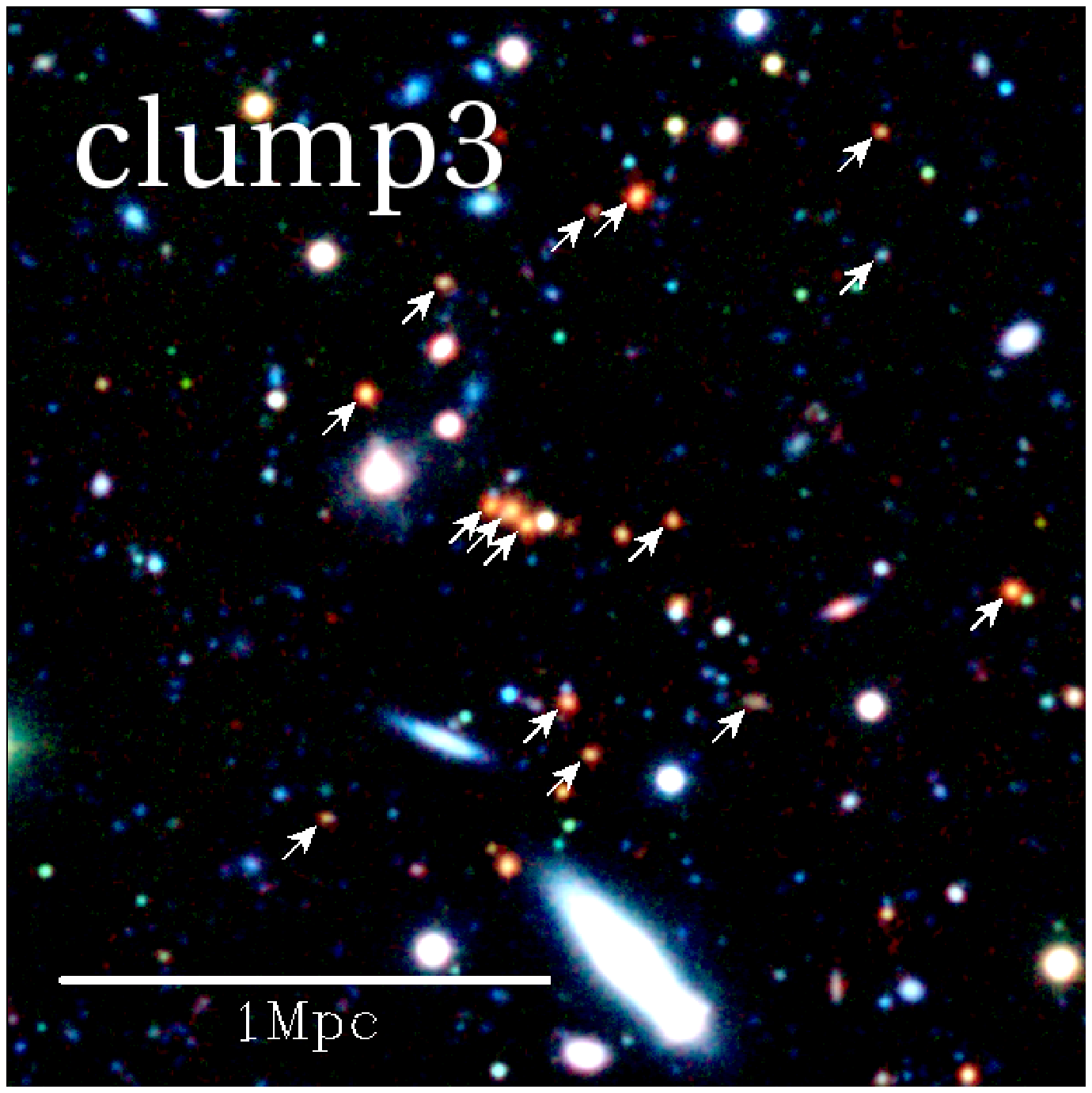}
\epsfxsize 0.33\hsize \epsfbox{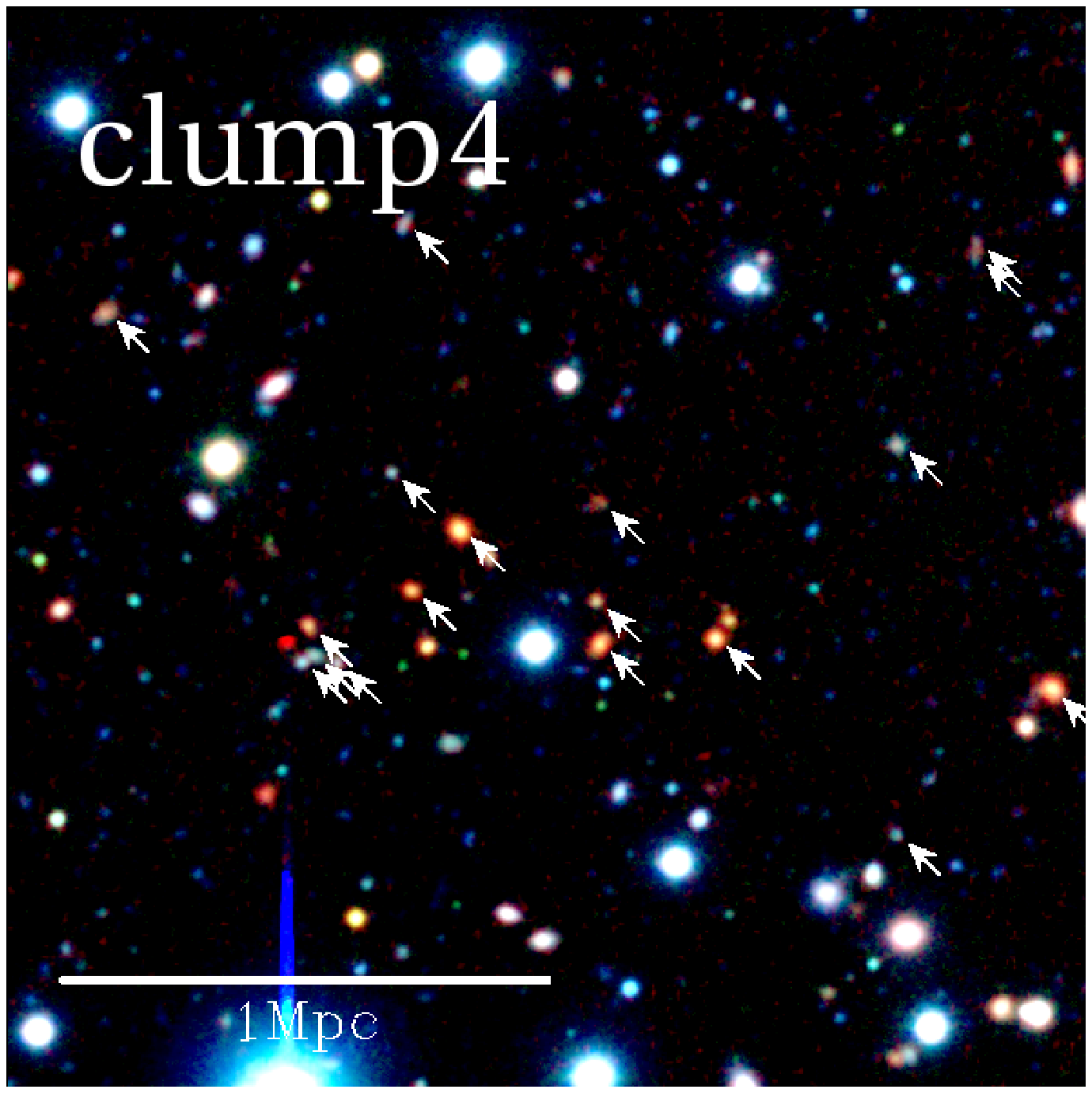}
\end{center}
\caption{
The 2 arcmin $\times$ 2 arcmin pseudo-colour images ($Rz'K$)
of the main cluster and four clumps.
The arrows indicate photo-$z$ selected galaxies ($1.0<z_{phot}<1.3$).
The 1 Mpc (comoving) scale is shown at bottom-left on each image.
North is up and East is to the left.
}
\label{fig:close_up}
\end{figure*}

We visually inspect all the significant over-density regions
and define four plausible clumps in Fig. \ref{fig:lss} dubbed clumps 1 to 4.
We show in Fig. \ref{fig:close_up} the close-up views of the main cluster and four clumps.
We find a clear concentration of red galaxies in each field except the clump 2.
The concentrations of the red galaxies in the three clumps 
suggest that they are physically bound galaxy systems at $z\sim1.2$.
The clump 2 is somewhat extended and is likely a filament. 
Its over-density of red galaxies is significant as shown in
Figs. \ref{fig:lss} and  \ref{fig:close_up}, although the red galaxies are
not strongly concentrated.
The colour of the red galaxies in the clumps is very similar to that of
the main cluster galaxies, suggesting that the clumps lie near the cluster redshift.

\subsection{Comparison with X-ray}

It is interesting to compare the observed structure with an X-ray map.
If the clumps are dynamically bound systems, we expect
diffuse X-ray emissions from hot thermal plasma trapped in the potential wells of the systems.
In other words, if the clumps are detected in X-ray, it is strong
evidence for bound systems.

Two observations performed by XMM-Newton could be used to search for extended
X-ray emission in the RDCSJ1252 field. Their observational IDs are 0057740301
and  0057740401 \citep{rosati04}. After the cleaning for background flares, the total exposure times
are 104ksec for pn, 128ksec MOS1 and 124 ksec in MOS2. A sophisticated
background removal technique has been applied, as described in \citet{finoguenov06}
and in addition 
we have removed the detected point sources on all spatial scales.

\begin{figure}
\begin{center}
\leavevmode
\epsfxsize 1.0\hsize \epsfbox{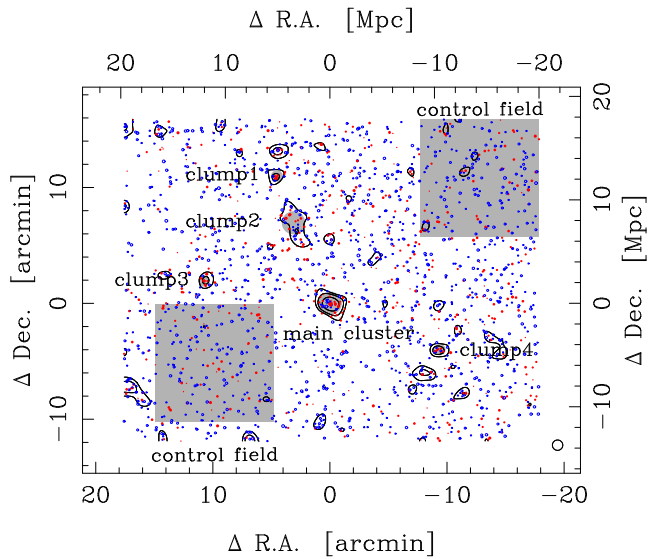}
\end{center}
\caption{
Distribution of photo-$z$ selected ($1.0<z_{phot}<1.3$)
galaxies around RDCSJ1252.
Galaxies are separated into red and blue galaxies at $i'-K=2.5$
and are shown as filled and open circles, respectively.
The contours show the over-density regions at 3$\sigma$, 5$\sigma$ and 10$\sigma$
significance levels relative to the density fluctuation in the control field.
Galaxy densities are smoothed with a Gaussian kernel with a scale of 0.5 Mpc
(comoving) at the cluster redshift.
The circle at the bottom-right shows the smoothing scale.
The shaded regions are defined as clump1, 2, 3, 4, main cluster, and control field.
The top and right ticks show the comoving scales.
North is up and East is to the left.
}
\label{fig:lss}
\end{figure}

\begin{figure}
\begin{center}
\leavevmode
\epsfxsize 1.0\hsize \epsfbox{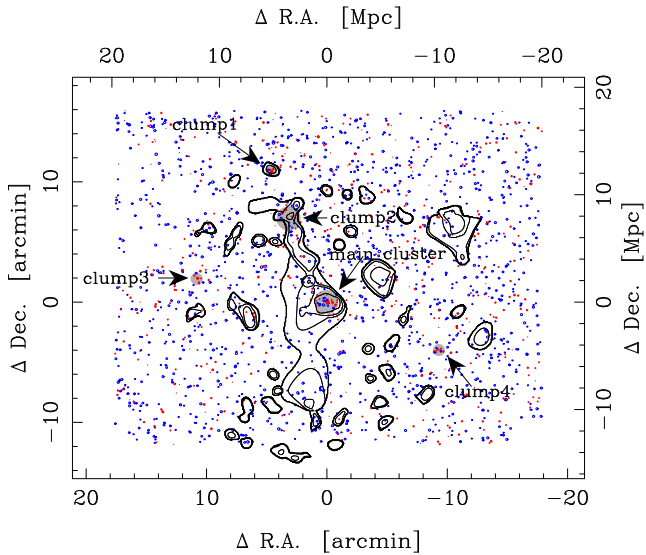}
\end{center}
\caption{
The distribution of photo-$z$ selected galaxies ($1.0<z_{phot}<1.3$)
with the X-ray contours overlaid.
The contours show the flux levels of
$1.0\times10^{-17}$,  $4.0\times10^{-17}$,  $1.6\times10^{-16}$,
$6.3\times10^{-16}$, $2.5\times10^{-15}$, and $1.0\times10^{-14}$ $\rm ergs\ s^{-1}\ cm^{-2}$.
}
\label{fig:xray}
\end{figure}

The resulting wavelet-filtered at minimal significance of 4$\sigma$ image of
the extended X-ray emission is shown as overlay in Fig. \ref{fig:xray}.
In addition to the emission from the main cluster,
extended emission is seen at the locations of the clumps 1 and 2
as well as emission associated with a number of foreground structures.
The extended X-ray emission around the clump 2 extends towards south
and traces the galaxy distribution (see the elongated galaxy distribution
in Fig. \ref{fig:lss}).
The detection of the clump2 may be surprising given the somewhat
extended galaxy distribution.
Interestingly, we observe no foreground structures between the clump 2
and main cluster, and the extended emission might suggest
an early merger of a few groups around clump 2.

We perform an estimate of the significance of the extended X-ray emission,
associated with the main cluster and all the clumps over the contribution of
point sources.
We find that clumps 1 and 2 are detected at $5.9\sigma$ and $4.5\sigma$,
respectively.
Clump 3 is marginally detected at the significance of $2.4\sigma$
and clump 4 is not detected.
The flux of the main cluster is measured at $25\sigma$.
Note that the southern extension of the clump 2 is detected at $5.4\sigma$.

Table \ref{tab:xray} shows the results of flux estimates and
corresponding mass estimates for the main cluster and the clumps
based on scaling relations as described in \citet{finoguenov06}.
We assume that all the clumps lie at z=1.24 and quote the 3 sigma upper limit for clump 4.
We also present the results for the southern extension of the clump 2.
The error bars on mass estimate are formal, while the precision of mass estimate is
about 40\% and is driven by uncertainty in the evolution in the scaling
relations as well as the scatter in the $\rm L_X-M$ relation.
Note that the main cluster has $M_{500}=1.5\times10^{14}\rm M_\odot$, which is slightly
smaller than that reported in \citet{rosati04} but within the 40\% uncertainty.
Note as well that we see a possible foreground system at the northern part of clump 2,
which may result in the corresponding overestimate of the mass of the structure.
The discovered clumps turn out to be lower mass systems by a factor of $>3$.
Clump 4 is not detected at a significant level, but it does not necessarily
mean that clump 4 is not a bound system.  It may be a very low mass system.

As we have demonstrated here, some of the clumps show significant X-ray emissions.
Together with the strong concentrations of red galaxies,
it is strongly suggested that these clumps are bound systems.
However, it is still unclear if they actually lie at the cluster redshift
or at slightly different redshifts, and
spectroscopic redshifts are needed to confirm the structure.

\begin{table*}
\begin{center}
\caption{X-ray properties of the RDCS1252 field.  All the clumps are assumed to be at $z=1.24$.
The clump 4 is not detected in X-ray and the values quoted below are $3\sigma$ limits.}
\label{tab:xray}
\begin{tabular}{ccccccc}
\hline
 & R.A.    & Dec.    &$r_{500}$&  M$_{500}$   & L$_{\rm 0.1-2.4 keV}$ & flux\\
 & (J2000) & (J2000) & (arcmin)& ($10^{13}$ M$_\odot$) &  ($10^{42}$ ergs s$^{-1}$) &  ($10^{-15}$ ergs s$^{-1}$ cm$^{-2}$)\\
\hline
 Clump 1& $12^h 53^m 15^s.5$ & $-29^\circ 16' 24''$ &  0.57   & $3.7\pm0.6$  & $48\pm8$  & $3.0\pm0.5$\\
 Clump 2& $12^h 53^m 08^s.8$ & $-29^\circ 20' 06''$ &  0.56   & $3.6\pm0.7$  & $46\pm10$ & $2.7\pm0.6$\\
 Clump 2 - South & $12^h 53^m 06^s.7$ & $-29^\circ 21' 35''$ & 0.54& $3.2\pm0.5$ & $40\pm7$  & $2.4\pm0.4$\\
 Clump 3& $12^h 53^m 43^s.9$ & $-29^\circ 25' 18''$ &  0.43   & $1.6\pm0.6$  & $19\pm8$  & $0.8\pm0.3$\\
 Clump 4& $12^h 52^m 11^s.4$ & $-29^\circ 31' 14''$ & $<0.45$ & $<1.9$       & $<23$     & $<1.1$\\
Main Cluster& $12^h 52^m 55^s.1$ & $-29^\circ 27' 23''$ &  0.90   & $14.7\pm0.5$ & $220\pm8$ & $188\pm0.7$\\

\hline
\end{tabular}
\end{center}
\end{table*}

%
%
\section{NTT OBSERVATION}
\label{sec:cmd}

Following the discovery of the large scale structure surrounding
RDCSJ1252, we used SOFI \citep{moorwood98} on the ESO NTT
to obtain deeper images of the clumps 1 to 4 in the $K_s$ filter. We used large
field imaging mode of SOFI, which results in a field of view of 5 arcmin $\times$ 5 arcmin
and a pixel scale of 0.29" per pixel. The observing dates, the mean
exposure time and the depth of the images are reported in Table \ref{tab:obs}.
Photometric standards from the list of \citet{persson98} were taken at
frequent intervals during the observations. Over the three night of
observation, the RMS in the zero point (10 standards) was 0.02 magnitudes.

The SOFI data were reduced in a standard manner: the zero level offset
was removed by subtracting a dark frame, the pixel-to-pixel
sensitivity variations were removed with dome flats, and the sky was removed by
subtracting a sky frame that is estimated from the data. We used the
{\sc xdimsum} package in {\sc iraf} to remove the sky. The sky subtracted frames were then
registered and combined.
As a final step, we remove artifacts that come from bright stars and residual variations in
the zero level offset. The artifacts are constant along detector rows, so an effective way of
removing them is to collapse (a clipped average) the 2 dimensional data along rows and
to subtract the resulting clipped 1 dimensional vector from the 2 dimensional image.

The FWHM of stars on the resulting images averages 0.6 arcsec. To match the seeing
in the Suprime-Cam images ($\sim0.8$ arcsec), we convolve the SOFI data with a 
2 dimensional Gaussian. We then construct a $K_s$ selected catalogue using
{\sc sextractor}. We adopt MAG\_AUTO for total magnitudes and derive the colours
in 1.5 arcsec diameter apertures.
Note that we do not use total magnitudes for colors in what follows.

We also use the $K_s$-band SOFI image of the main cluster presented in \citet{lidman04}.
Details of the observation is described in \citet{lidman04}.
The magnitude limit is shallower by 0.4 magnitude than the other four images
due to the shorter exposure time ($90$ min).
Note that the limiting magnitudes quoted for all the SOFI images are
evaluated in the same manner as for the Suprime-Cam and WFCAM images.
The image is smoothed to the Suprime-Cam seeing size.
The photometric catalogue is constructed in the same way as in the four fields.

We compare SOFI photometry with WFCAM one using common stellar objects
and find a reasonable agreement; 
the median of $K_{WFCAM}-K_{s,\ SOFI}$ and dispersion around it are
$+0.04$ and $0.03$ at $K_{s, SOFI}<19$.
A comparison between optical-nearIR colours derived from total magnitudes
(the ones we have used in the previous section) and those from aperture magnitudes
(the ones we use in what follows) shows a $\lesssim0.05$ mag. offset
with a $\sim0.05$ mag. scatter.
The scatter is comparable to the photometric errors and
it turns out that the WFCAM image gives reasonably good photometry.
The observed systematic offsets are likely the primary cause of the offset
in the photometric redshifts presented in the previous section.
We rerun the photo-$z$ code with the aperture magnitudes and find
that the photo-$z$ offset is reduced to $z_{phot}-z_{spec}=-0.02$
for galaxies at $1.2<z_{spec}<1.3$.

%
%
\section{Colour-Magnitude Diagrams}
\label{sec:cmd}

\subsection{Redshifts of the Clumps}

Following the discovery of the four clumps,
we examine the redshifts of the individual clumps in more detail.
A quick and effective way of obtaining redshifts of galaxy systems is
to see the location of red galaxy sequence on the colour-magnitude diagram
(CMD; e.g. \citealt{bower92,gladders00}).
Note that spectroscopic data are available only for the central
cluster (Demarco et al. submitted.), and no spectroscopic data are currently
available for the newly found clumps. 
We adopt a 60 arcsec diameter aperture for the clumps 1, 3 and  4, which
is equivalent to $500\rm\ h_{70}^{-1}\ kpc$ (physical), to define the clumps.
For the clump 2, we adopt a 120 arcsec aperture ($1\rm\ h_{70}^{-1}\ Mpc$ )
because of its relatively large spatial extent.
Galaxies are extracted from these apertures for the analysis below.

We present in Fig. \ref{fig:cmd} the CMDs of the main cluster, clumps 1, 2, 3 and 4.
All the galaxies in the five SOFI fields are also shown for reference.
Here we do not apply the photometric redshift selection nor the conventional
statistical contamination subtraction on the CMD.
Galaxies in the main cluster exhibit a clear colour-magnitude relation (CMR).
We perform a biweight fit to the relation excluding the obvious outliers.
Galaxies with $2.5<i'-K_s<3.5$ shown as the dotted lines in Fig. \ref{fig:cmd} are used in the fit.
The best-fitting line is $i'-K_s=(-0.058^{+0.026}_{-0.032})\times(K-20)+(3.13^{+0.04}_{-0.04})$
and is shown by the solid line.
The errors are estimated from bootstrap resampling of the input galaxies
taking into account the photometric errors.
An important test here is whether the CMRs of
the clumps are offset from the main cluster relation.
An offset to the bluer side suggests that a clump lies at a slightly lower
redshift, while an offset to the redder side suggests a slightly higher redshift.

In each clump, we select $2.5<i'-K_s<3.5$ galaxies and fit a CMR
by offsetting the main cluster relation with its slope fixed\footnote{
The poor statistics does not allow us to examine the variation
in the slopes in the clumps.
We note however that if we fit the CMR allowing the slope to vary,
the best-fitting slope is consistent with the main cluster one within its somewhat large error.
}.
Our measurements are summarized in Table \ref{tab:cmr}.
The best-fitting CMRs are shown as the dashed lines in Fig. \ref{fig:cmd}.
We find that CMRs in the clumps 1 and 4 are consistent with the main cluster
relation within the error, suggesting that their redshifts are very close
to the cluster redshift.
On the other hand, the clumps 2 and 3 show slightly bluer CMRs
[$\Delta(i'-K_s)\sim-0.1$ to $-0.2$].
We note that an $i'-K$ colour difference of $-0.10$ mag around the cluster redshift
translates into a redshift difference of $-0.12$ for passively evolving galaxies
formed at $z_f=2.5$\footnote{
If we attribute the $i'-K$ colour difference of $-0.1$ mag. to
age differences assuming that the clumps lie at the same redshift,
their formation epochs differ by $\sim +0.3$ Gyr.
}.
The clumps 2 and 3 might lie at slightly lower redshifts.
However, we need spectroscopic redshifts to prove or disprove the physical association
between the discovered clumps and the main cluster.
In any case, it is likely that these clumps are located near the cluster redshift,
and it is reasonable to combine these clumps to investigate the properties of galaxy groups
at $z\sim1.2$ (see the next subsection).

\subsection{The Deficit of Faint Red Galaxies}

Recent work on the CMR suggested that there is a lack of faint red galaxies
in high redshift clusters compared to in local clusters
\citep{kajisawa00,nakata01,kodama04,delucia04,tanaka05,delucia06,homeier06}.
An emerging picture is that the CMR first appears at bright magnitudes and
it extends to faint magnitudes with time.
Following these studies, we examine the faint end of the CMR in detail.

\begin{figure}
\begin{center}
\leavevmode
\epsfxsize 1.0\hsize \epsfbox{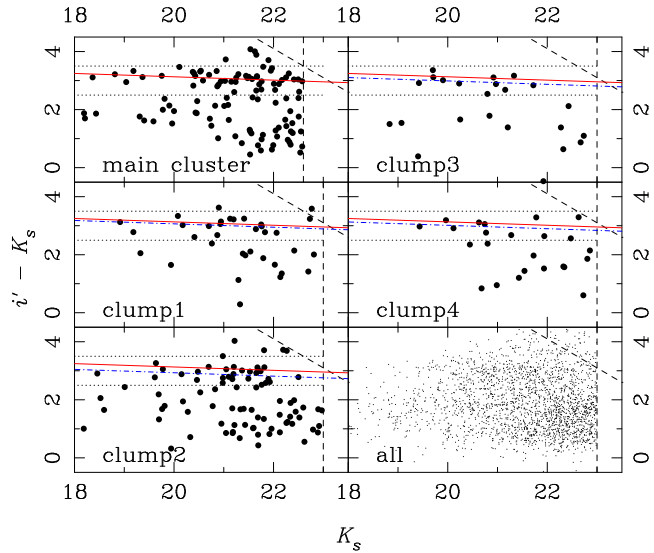}
\end{center}
\caption{
The $i'-K_s$ versus $K_s$ CMD of the main cluster and clumps 1 to 4.
All the galaxies in the five SOFI fields are also plotted in
the bottom-right panel.
In each panel, the dashed lines indicate the $5\sigma$ limiting magnitudes and colours.
The solid red line is the CMR determined
from the main cluster sample.
The best-fitting CMR for each clump is determined
from galaxies within the dotted lines ($2.5<i'-K<3.5$)
and is shown as the dot-dashed line (see also Table \ref{tab:cmr}).
}
\label{fig:cmd}
\end{figure}

\begin{figure}
\begin{center}
\leavevmode
\epsfxsize 1.0\hsize \epsfbox{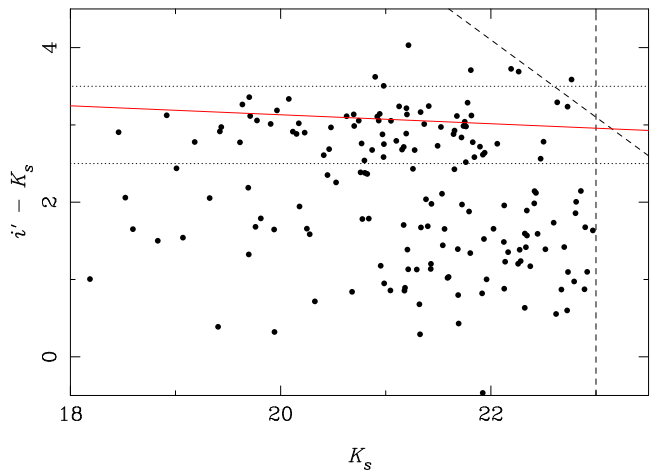}
\end{center}
\caption{
The CMD ($i'-K_s$ vs. $K_s$) of the composite clump.
The dashed lines show the $5\sigma$ limiting magnitudes and colours.
The dotted lines show $i'-K=2.5$ and 3.5, and the solid
lines is the best-fitting CMR for the main cluster.
}
\label{fig:composite_cmd}
\end{figure}

The CMDs of the clumps shown in Fig. \ref{fig:cmd} seem to suggest that
the CMR is truncated at $K_s=22$.  Few faint red galaxies can be found on the CMR.
This trend is particularly noticeable in the clump 2.
To gain the statistics and clarify the possible deficit of faint red galaxies in the clumps,
we make a composite clump from the clumps 1 to 4 and plot a composite CMD in Fig. \ref{fig:composite_cmd}.
It is important to repeat that we do not apply the photometric redshift selection nor
conventional statistical field subtraction.
Therefore, we are free from any bias and uncertainty arising from
the contamination subtraction.

Fig. \ref{fig:composite_cmd} is the highlight of this paper ---
we observe the CMR with a very clear deficit of faint red galaxies at $K_s>22$.
The CMR is very sharply truncated at $K_s=22$ and only a few faint red galaxies populate in the clumps.
This truncation magnitude is consistent with those reported in \citet{kajisawa00}
and \citet{nakata01}, although their results are not as convincing as we see here.
Interestingly, no such sharp truncation of the CMR is seen in the main cluster (see Fig. \ref{fig:cmd}).
We observe a clear CMR down to $K=22.6$.
\citet{lidman04} carried out deep ISAAC imaging and showed that the CMR extends
to magnitude as faint as $K_s=23$ ($K_s\sim21$ on the Vega system).

One may suspect that the detection of faint objects fails in dense
environments due to crowdedness and it causes the apparent lack of faint red galaxies.
However, the deficit is clearly seen even in the somewhat loose clump 2.
Also, a clear CMR is observed below $K_s=22$ in the main cluster,
which is the densest environment in our sample.

Therefore, the observed deficit is not due to blending effects.

\begin{figure}
\begin{center}
\leavevmode
\epsfxsize 1.0\hsize \epsfbox{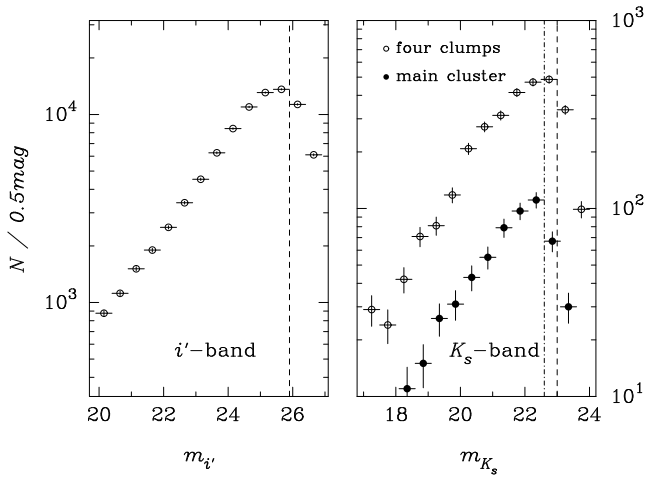}
\end{center}
\caption{
The number distributions of Suprime-Cam $i'$--detected (left) and
SOFI $K_s$--detected (right) objects in 0.5 mag. bins.
The vertical error bars show the Poisson errors and the horizontal bars the size of the bin.
The vertical line is the $5\sigma$ magnitude limit.
In the right panel, the points for the four clumps (i.e., our data)
and for the main cluster (i.e., data from \citealt{lidman04}) are plotted separately
since the limiting magnitudes are different.
The dashed and dot-dashed vertical lines in the right panel show the limiting magnitudes
for the four clumps and the main cluster, respectively.
}
\label{fig:completeness}
\end{figure}

The magnitude $K_s=22$ is somewhat close to the magnitude limit, but
the lack of faint red galaxies is not due to incompleteness effects either.
We present in Fig. \ref{fig:completeness} the number distribution of $i'$-
and $K_s$-detected objects as a function of magnitude.
The numbers of objects increase towards faint magnitudes and then
turns to decrease near the magnitude limits, suggesting that
incompleteness effects come in to play there.
As a rough estimate of completeness of the composite clump,
we fit a power law to the galaxy distribution at $20<i'<24$ and $17<K_s<22$
(for the composite clump), extrapolate it to the magnitude limits,
and compare the expected number of objects with the observed counts.
We find that we are $\sim70$ per cent complete at the magnitude limit.
We are more than 80 per cent complete at $22<K_s<22.5$,
at which the CMR is truncated.  It is very unlikely that
this level of incompleteness causes the sharp truncation of
the CMR at $K_s=22$ in the composite clump.
As shown below, the number of faint red galaxies at $K_s>22$ in the composite clump
is smaller by nearly an order of magnitude than that in the main cluster
and the observed deficit is not due to such small incompleteness effects.
Therefore, we argue that the observed deficit is real.

\begin{figure}
\begin{center}
\leavevmode
\epsfxsize 1.0\hsize \epsfbox{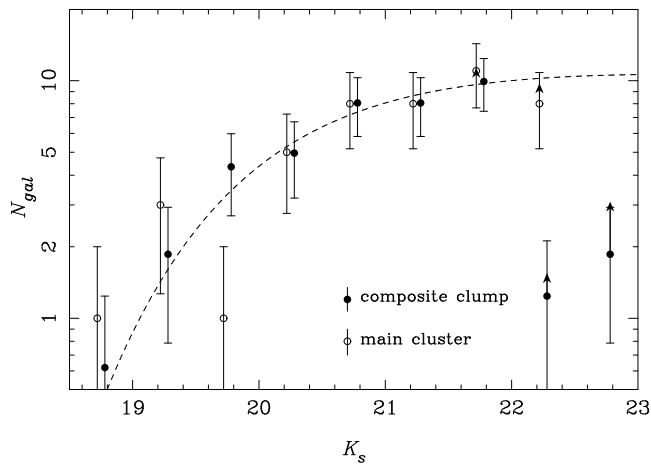}
\end{center}
\caption{
LFs of the red galaxies in the composite clump (filled circles)
and main cluster (open circles).
Since the magnitude limit of the main cluster field is $K_s=22.6$, we cannot draw
a point at $22.5<K_s<23.0$.
The arrows indicate the corrections for incompleteness at the two faintest magnitude bins
for each sample based on the $N-m$ relations presented in Fig. \ref{fig:completeness}.
The data points for the composite clump are scaled to match the main cluster LF at $K_s\sim21$.
The dashed line shows the Schechter function from \citet{strazzullo06},
which is scaled to fit the observed points for an illustrative purpose.
}
\label{fig:lf}
\end{figure}

We quantify the deficit with luminosity functions (LFs) of red galaxies.
We again extract galaxies having $2.5<i'-K_s<3.5$, which is broad enough to
include galaxies on the CMR, and plot their LFs in Fig. \ref{fig:lf}.
No field correction has been applied.
The deficit of faint red galaxies in the composite clump is apparent.
The number of red galaxies sharply decreases at $K_s>22$ and the LF
shows a 'break' at $K_s=22$.
No such sharp decline is observed in the main cluster.
We plot the best-fit Schechter function of $K_s$ selected galaxies
in the main cluster from \citet{strazzullo06} as the dashed line.
Their LF includes blue galaxies as well as red ones, but the cluster
is dominated by red galaxies and we may use the LF as a proxy for the LF of red galaxies.
In fact, the LF gives a good fit to the observed red LF (open circle).
The LF is nearly flat at the faint end and no sharp decline
is seen at faint magnitudes.
This is again in contrast with the sharp decline seen in the composite clump.

To sum up, we observe the clear deficit of faint red galaxies in the clumps.
The CMR is sharply truncated at $K_s=22$ and few faint red galaxies are seen.
Interestingly, no such sharp truncation is observed in the main cluster.
We discuss implications of these findings have for the galaxy evolution
in the next section.

\begin{table}
\begin{center}
\caption{
The CMRs in the main cluster and the discovered clumps.
The second column shows $i'-K$ colour of the CMR at $K=20$.
}
\label{tab:cmr}
\begin{tabular}{ccc}
\hline
Region               & $(i'-K)_{K=20}$ \\
\hline
{\it Main Cluster}   & $3.13^{+0.04}_{-0.04}$   \\
{\it Clump1}         & $3.06^{+0.05}_{-0.06}$   \\
{\it Clump2}         & $2.93^{+0.04}_{-0.04}$   \\
{\it Clump3}         & $2.99^{+0.06}_{-0.07}$   \\
{\it Clump4}         & $3.01^{+0.08}_{-0.08}$   \\
\hline
\end{tabular}
\end{center}
\end{table}

%
%
\section{Discussion}
\label{sec:discussion}

Figs. \ref{fig:composite_cmd} and \ref{fig:lf} provide strong evidence for 'down-sizing'
in star formation activities of galaxies \citep{cowie96}.
The population of galaxies that host active star formation activities is
progressively shifted to lower mass galaxies with time.
This drives the build-up of the CMR from the bright end to the faint end.
The faint end of the CMRs in the clumps is still under construction at $z\sim1.2$,
and will be built-up later at lower redshifts as observed in \citet{tanaka05}.

We also find that the CMR build-up depends on environment.
The poor (i.e., low-mass) clumps show the sharp truncation of the CMR, while
the main cluster does not.
Therefore, the CMR build-up is 'delayed' in poor systems.
This is consistent with our previous finding
(\citealt{tanaka05}; but see also \citealt{delucia06}).
All these results lead us to suggest that galaxies follow
the 'environment-dependent down-sizing' evolution.
Massive galaxies in dense environments stop forming stars at early times and become red.
Less massive galaxies in less dense environments continue
to form stars at later times and they get red at lower redshifts.
It is interesting to note that the main cluster itself shows
a hint of the deficit of faint red galaxies compared to local clusters
(\citealt{toft04}; but see also \citealt{strazzullo06}).
The faint end of the main cluster CMR is possibly still under construction,
although the clear CMR is already established.

It is interesting to examine how the magnitude at which the CMR is truncated changes with time.
\citet{tanaka05} found that the CMR disappears at $M_V\sim-20$ in $z\sim0.8$ groups.
They also found that galaxies in $z\sim0.5$ groups form a tight CMR down to $M_V=-18$
and it possibly extends to fainter magnitudes.
Assuming that the clumps we discover are the progenitors of the $z\sim0.5$ and 0.8 groups,
the passive evolution model ($z_f=2.5$) predicts that the galaxies with $M_V=-20$ at $z=0.8$
and those with $M_V=-18$ at $z=0.5$ would appear $K_s=22.4$ and $K_s=24.1$ at $z=1.2$,
respectively.
We find in this paper that the CMR in the clumps at $z=1.2$ is sharply truncated
at $K_s=22$ (see Fig. \ref{fig:composite_cmd}).
Therefore, it seems that the truncation magnitude gets fainter at lower redshifts.

We linearly fit the brightening of the truncation magnitude
as a function of look-back time over $0.5<z<1.2$
(we assume that the $z\sim0.5$ CMR is truncated at $M_V=-18$).
If we could extrapolate this relation to higher redshifts,
we would see further brightening in the truncation magnitude by 2 mags.
at $z\sim2.5$, meaning that the CMR would eventually
disappear even at the brightest-end.
This epoch ($z\sim2.5$) coincides with the formation redshift of
the red galaxies in the main cluster inferred from the CMR \citep{lidman04}.

We admit that the assumptions we have made here are not necessarily valid.
For example, we have assumed that the $z\sim0.5$ CMR is truncated at $M_V=-18$,
but it may extends to fainter magnitudes.
The predicted redshift $z\sim2.5$ at which the brightest end of the CMR appears
should be regarded as an upper limit --- if the $z\sim0.5$ CMR extends
fainter than $M_V=-18$, then the truncation magnitude gets fainter more rapidly 
with time and the CMR would appear at lower redshifts than $z=2.5$.
Another concern is that galaxies in the $z\sim1.2$ clumps may not be the progenitors of
those seen in the $z\sim0.8$ and 0.5 groups and therefore they may not lie
on the same evolutionary track.
We should somehow establish a robust evolutionary link of galaxies
in different environments at different redshifts in a statistical way
to further discuss the evolution of the truncation magnitude
and reach a firm conclusion.
This will eventually be achieved with intensive spectroscopic
observations with the coming near-IR facilities (e.g., FMOS; \citealt{kimura03})
and with a help of numerical simulations that predict the assembly history
of clusters (e.g. Millennium Run; \citealt{springel05}).

%
%
\section{Summary}
\label{sec:summary}

We carried out wide-field multi-band imaging of RDCSJ1252 at $z=1.24$.
Taking advantage of the multi-band data, we apply the photometric redshift technique
to reduce the foreground and background contamination.
The distribution of $z\sim1.2$ galaxies reveals clumpy structures
surrounding the central cluster.
We see a chain of clumps towards the NNE direction extending over 15 Mpc.
There are a few more clumps located at ENE and WSW of the main cluster.
No such clumps are seen at NW and SE of the main cluster.
It seems there is a large ($>20$ Mpc) filamentary structure extending in the NE-SW direction.
We compare the galaxy distribution with deep X-ray imaging and find that
some of the clumps show significant X-ray emissions.
Together with the concentrations of red galaxies in the clumps,
it is strongly suggested that they are physically bound systems.

Following the discovery of the possible large-scale structure,
we carried out the deep SOFI imaging of the four plausible clumps.
A close inspection of the CMR of red galaxies in the clumps revealed
that some of them may lie at slightly lower redshift
than the cluster redshift ($\Delta z\sim-0.1$).
However, because of the limited accuracy of the redshifts derived from CMRs,
spectroscopic follow-up observations are essential to confirm
the physical clustering of the individual clumps and their physical 
association to the main body of the cluster.

We make the composite CMD of the four clumps.
We find that the CMR is sharply truncated at $K_s=22$.
Few faint red galaxies reside in the clumps.
The deficit of faint red galaxies is quantified with the LF of the red galaxies.
The LF shows a prominent 'break' at $K_s=22$ and the number of red galaxies
sharply decreases at fainter magnitudes.
In contrast, the main cluster shows a clear CMR down to fainter magnitude of $K=23$ \citep{lidman04}
and its red LF does not show any deficit of faint red galaxies.
From these findings and by comparing with lower redshift systems, we suggest that
(1) the CMR first appears at bright magnitudes and extends to fainter magnitude at later times,
and (2) the build-up of the CMR is delayed in lower-mass systems.
We therefore suggest that galaxies follow the 'environment-dependent down-sizing' evolution.
Massive galaxies in high density environments first stop forming stars and get red.
Less massive galaxies in less dense environments become red at lower redshifts.
Based on a few assumptions, we predict that the first appearance of the CMR in
galaxy groups will be observed at $z\sim2.5$.

%
%
\section*{Acknowledgements}
We thank the anonymous referee for the careful reading of the paper
and for helpful suggestions, which significantly improved the paper.
We thank Naoyuki Tamura for helpful comments.
M.T. acknowledges support from the Japan Society for Promotion of Science (JSPS)
through JSPS research fellowships for Young Scientists.
This work was financially supported in part by a Grant-in-Aid for the
Scientific Research (Nos.\, 15740126 and 18684004) by the Japanese Ministry of Education,
Culture, Sports and Science.
T.K. acknowledges hospitarity of ESO during his stay in 2005 when a part of this
work was done.
This study is based on data collected at Subaru Telescope, which is operated by
the National Astronomical Observatory of Japan, and on
observations made with the United Kingdom Infrared Telescope, which 
is operated by the Joint Astronomy Centre on behalf of the U.K.
Particle Physics and Astronomy Research Council.
The use of the UKIRT 3.8-m telescope for the observation is supported by NAOJ.
This study is also in part based on data collected at New Technology
Telescope at the ESO La Silla Observatory.
In Germany, the XMM--Newton project is supported by the Bundesministerium
fuer Wirtschaft und Technologie/Deutsches Zentrum fuer Luft- und Raumfahrt
(BMWI/DLR, FKZ 50 OX 0001), the Max-Planck Society and the
Heidenhain-Stiftung. Part of this work was supported by the Deutsches
Zentrum f\"ur Luft-- und Raumfahrt, DLR project numbers 50 OR 0207 and 50 OR
0405.  AF acknowledges support from BMBF/DLR under
grant 50 OR 0207, MPG.

%
%


\begin{thebibliography}{99}
\bibitem[\protect\citeauthoryear{Bertin \& Arnouts}{1996}]{bertin96}
Bertin E., Arnouts S., 1996, A\&AS, 117, 393
\bibitem[\protect\citeauthoryear{Blakeslee et al.}{2003}]{blakeslee03}
Blakeslee J.~P., et al., 2003, ApJ, 596, L143
\bibitem[\protect\citeauthoryear{Bolzonella, Miralles, \& Pell{\'o}}{2000}]{bolzonella00}
Bolzonella M., Miralles J.-M., Pell{\'o} R., 2000, A\&A, 363, 476 
\bibitem[\protect\citeauthoryear{Bower, Lucey, \& Ellis}{1992}]{bower92}
Bower R.~G., Lucey J.~R., Ellis R.~S., 1992, MNRAS, 254, 601
\bibitem[\protect\citeauthoryear{Cowie et al.}{1996}]{cowie96} 
Cowie L.~L., Songaila A., Hu E.~M., Cohen J.~G., 1996, AJ, 112, 839
\bibitem[\protect\citeauthoryear{De Lucia et al.}{2004}]{delucia04}
De Lucia G., et al., 2004, ApJ, 610, L77 
\bibitem[\protect\citeauthoryear{De Lucia et al.}{2006}]{delucia06}
De Lucia G., et al., 2006, astro, arXiv:astro-ph/0610373
\bibitem[\protect\citeauthoryear{Demarco et al.}{2005}]{demarco05}
Demarco R., et al., 2005, A\&A, 432, 381
\bibitem[\protect\citeauthoryear{Finoguenov et al.}{2006}]{finoguenov06}
Finoguenov A., et al., 2006, astro, arXiv:astro-ph/0612360 
\bibitem[\protect\citeauthoryear{Gal \& Lubin}{2004}]{gal04} 
Gal R.~R., Lubin L.~M., 2004, ApJ, 607, L1
\bibitem[\protect\citeauthoryear{Girardi et al.}{2005}]{girardi05}
Girardi M., Demarco R., Rosati P., Borgani S., 2005, A\&A, 442, 29 
\bibitem[\protect\citeauthoryear{Gladders \& Yee}{2000}]{gladders00}
Gladders M.~D., Yee H.~K.~C., 2000, AJ, 120, 2148 
\bibitem[\protect\citeauthoryear{Gunn \& Stryker}{1983}]{gunn83}
Gunn J.~E., Stryker L.~L., 1983, ApJS, 52, 121
\bibitem[\protect\citeauthoryear{Henry et al.}{2000}]{henry00} 
Henry D.~M., Atad-Ettedgui E., Casali M.~M., Bennett R.~J., Bridger A., Ives D.~J., Rae R.~G., Hawarden T.~G., 2000, SPIE, 4008, 1325 
\bibitem[\protect\citeauthoryear{Holden et al.}{2005}]{holden05}
Holden B.~P., et al., 2005, ApJ, 620, L83
\bibitem[\protect\citeauthoryear{Homeier et al.}{2006}]{homeier06}
Homeier N.~L., et al., 2006, ApJ, 647, 256 
\bibitem[\protect\citeauthoryear{Jarrett et al.}{2000}]{jarrett00}
Jarrett T.~H., Chester T., Cutri R., Schneider S., Skrutskie M., Huchra J.~P., 2000, AJ, 119, 2498 
\bibitem[\protect\citeauthoryear{Kajisawa et al.}{2000}]{kajisawa00}
Kajisawa M., et al., 2000, PASJ, 52, 61 
Kimura M., et al., 2003, SPIE, 4841, 974 
\bibitem[\protect\citeauthoryear{Kodama, Bell, \& Bower}{1999}]{kodama99}
Kodama T., Bell E.~F., Bower R.~G., 1999, MNRAS, 302, 152 
\bibitem[\protect\citeauthoryear{Kodama et al.}{2001}]{kodama01} 
Kodama T., Smail I., Nakata F., Okamura S., Bower R.~G., 2001, ApJ, 562, L9
\bibitem[\protect\citeauthoryear{Kodama et al.}{2004}]{kodama04} 
Kodama T., et al., 2004, MNRAS, 350, 1005 
\bibitem[\protect\citeauthoryear{Kodama et al.}{2005}]{kodama05} 
Kodama T., et al., 2005, PASJ, 57, 309 
\bibitem[\protect\citeauthoryear{Lidman et al.}{2004}]{lidman04} 
Lidman C., Rosati P., Demarco R., Nonino M., Mainieri V., Stanford S.~A., Toft S., 2004, A\&A, 416, 829 
 \bibitem[\protect\citeauthoryear{Lombardi et al.}{2005}]{lombardi05}
Lombardi M., et al., 2005, ApJ, 623, 42
\bibitem[\protect\citeauthoryear{Maughan et al.}{2006}]{maughan06}
Maughan B.~J., Ellis S.~C., Jones L.~R., Mason K.~O., C{\'o}rdova F.~A., Priedhorsky W., 2006, ApJ, 640, 219 
\bibitem[\protect\citeauthoryear{Miyazaki et al.}{2002}]{miyazaki02}
Miyazaki S., et al., 2002, PASJ, 54, 833
\bibitem[\protect\citeauthoryear{Moorwood, Cuby, \&  Lidman}{1998}]{moorwood98}
Moorwood, A., Cuby, J.G., Lidman, C. 1998, Messenger, 91, 9
\bibitem[\protect\citeauthoryear{Nakata et al.}{2001}]{nakata01}
Nakata F., et al., 2001, PASJ, 53, 1139 
\bibitem[\protect\citeauthoryear{Persson et al.}{1998}]{persson98}
Persson S.~E., Murphy D.~C., Krzeminski W., Roth M., Rieke M.~J., 1998, AJ, 116, 2475 
\bibitem[\protect\citeauthoryear{Rosati et al.}{2004}]{rosati04} 
Rosati P., et al., 2004, AJ, 127, 230
\bibitem[\protect\citeauthoryear{Schlegel, Finkbeiner, \& Davis}{1998}]{schlegel98}
Schlegel D.~J., Finkbeiner D.~P., Davis M., 1998, ApJ, 500, 525
\bibitem[\protect\citeauthoryear{Springel et al.}{2005}]{springel05}
Springel V., et al., 2005, Natur, 435, 629 
\bibitem[\protect\citeauthoryear{Strazzullo et al.}{2006}]{strazzullo06}
Strazzullo V., et al., 2006, A\&A, 450, 909 
\bibitem[\protect\citeauthoryear{Tanaka et al.}{2005}]{tanaka05}
Tanaka M., Kodama T., Arimoto N., Okamura S., Umetsu K., Shimasaku K., Tanaka I., Yamada T., 2005, MNRAS, 362, 268
\bibitem[\protect\citeauthoryear{Tanaka et al.}{2006}]{tanaka06}
Tanaka M., Kodama T., Arimoto N., Tanaka I., 2006, MNRAS, 365, 1392
\bibitem[\protect\citeauthoryear{Toft et al.}{2004}]{toft04} 
Toft S., Mainieri V., Rosati P., Lidman C., Demarco R., Nonino M., Stanford S.~A., 2004, A\&A, 422, 29 
\end{thebibliography}
\end{document}